\documentclass[prl,twocolumn,superscriptaddress,showpacs,preprintnumbers,amsmath,amssymb,floats]{revtex4}

\usepackage{txfonts}
\usepackage{amssymb}
\usepackage{graphicx}
\usepackage{amsmath}
\usepackage{amssymb}

\begin{document}

\title{Surface electronic structure and isotropic superconducting gap in (Li$_{0.8}$Fe$_{0.2}$)OHFeSe}

\author{X. H. Niu}
\author{R. Peng}
\author{H. C. Xu}
\author{Y. J. Yan}
\author{J. Jiang}
\author{D. F. Xu}
\author{T. L. Yu}
\author{Q. Song}
\author{Z. C. Huang}
\author{Y. X. Wang}
\author{B. P. Xie}
\affiliation{State Key Laboratory of Surface Physics, Department of Physics,  and Advanced Materials Laboratory, Fudan University, Shanghai 200433, People's Republic of China}
\affiliation{Collaborative Innovation Center of Advanced Microstructures, Nanjing 210093, People's Republic of China}
\author{X. F. Lu}
\author{N. Z. Wang}
\affiliation{Hefei National Laboratory for Physical Sciences at Microscale and Department of Physics, University of Science and Technology of China, Hefei, Anhui 230026, People's Republic of China}
\affiliation{Key Laboratory of Strongly-coupled Quantum Matter Physics, University of Science and Technology of China, Chinese Academy of Sciences, Hefei, Anhui 230026, People's Republic of China}
\author{X. H. Chen}
\affiliation{Hefei National Laboratory for Physical Sciences at Microscale and Department of Physics, University of Science and Technology of China, Hefei, Anhui 230026, People's Republic of China}
\affiliation{Key Laboratory of Strongly-coupled Quantum Matter Physics, University of Science and Technology of China, Chinese Academy of Sciences, Hefei, Anhui 230026, People's Republic of China}
\affiliation{Collaborative Innovation Center of Advanced Microstructures, Nanjing 210093, People's Republic of China}
\author{Z. Sun}\email{sunzhe@gmail.com}
\affiliation{National Synchrotron Radiation Laboratory, University of Science and Technology of China, Hefei, Anhui 230029, People's Republic of China}
\affiliation{Collaborative Innovation Center of Advanced Microstructures, Nanjing 210093, People's Republic of China}
\author{D. L. Feng}\email{dlfeng@fudan.edu.cn}
\affiliation{State Key Laboratory of Surface Physics, Department of Physics, and Advanced Materials Laboratory, Fudan University, Shanghai 200433, People's Republic of China}
\affiliation{Collaborative Innovation Center of Advanced Microstructures, Nanjing 210093, People's Republic of China}

\begin{abstract}

Using angle-resolved photoemission spectroscopy (ARPES), we revealed the surface electronic structure and superconducting gap of (Li$_{0.8}$Fe$_{0.2}$)OHFeSe, an intercalated FeSe-derived superconductor without antiferromagnetic phase or Fe-vacancy order in the FeSe layers, and with a superconducting transition temperature ($T_c$) $\sim$ 40 K. We found that (Li$_{0.8}$Fe$_{0.2}$)OH layers dope electrons into FeSe layers. The electronic structure of surface FeSe layers in (Li$_{0.8}$Fe$_{0.2}$)OHFeSe resembles that of Rb$_x$Fe$_{2-y}$Se$_2$ except that it only contains half of the carriers due to the polar surface, suggesting similar quasiparticle dynamics between bulk (Li$_{0.8}$Fe$_{0.2}$)OHFeSe and Rb$_x$Fe$_{2-y}$Se$_2$. Superconducting gap is clearly observed below $T_c$, with an isotropic distribution around the electron Fermi surface. Compared with $A_x$Fe$_{2-y}$Se$_2$ (\textit{A}=K, Rb, Cs, Tl/K), the higher $T_c$ in (Li$_{0.8}$Fe$_{0.2}$)OHFeSe might be attributed to higher homogeneity of FeSe layers or to some unknown roles played by the (Li$_{0.8}$Fe$_{0.2}$)OH layers.

\end{abstract}

%74.25.Jb: Electronic structure (photoemission, etc)
%74.70.Xa: Pnictides, Chalcogenides
%79.60.-i: Electron emission/photoemission
%71.20.-b: Electronic structure/condensed matter/crystalline solids

\maketitle

Heavily electron-doped iron selenide superconductors (HEDIS), such as   $A_x$Fe$_{2-y}$Se$_2$ (\textit{A}=K, Rb, Cs, Tl/K) \cite{XLChen, KFeSe40K1} and single-layer FeSe on oxides (SrTiO$_3$, BaTiO$_3$)  \cite{QKXue, XJZhou1, SYTan, XJZhou2, RPeng, RPengN}, are currently the research focus in the field of iron-based superconductors. The absence of hole Fermi surfaces  \cite{YZhangMat}, together with the nodeless superconducting gap \cite{YZhangMat,RPeng} in these materials pose great challenges on various pairing theories \cite{Hirschfeld}. Moreover, the dominant factors that induce high $T_c$ remains perplexing. In single-layer FeSe, evidence of record high $T_c$ of iron-based superconductors has been observed \cite{QKXue, XJZhou1, SYTan, XJZhou2, RPeng, RPengN, JFJia}. It is proposed that  the interaction with the interfacial oxide would significantly enhance the $T_c$ of FeSe \cite{RPengN,ZXShen}. However,  the air sensitivity of the FeSe films induces difficulties in systematic ex-situ characterizations, and thus the relationship between  $T_c$ and  interfacial interactions remains unsettled. In $A_x$Fe$_{2-y}$Se$_2$, there is an intrinsic mesoscopic phase separation of a superconducting phase and an Fe-vacancy ordered antiferromagnetc insulating phase \cite{Minghu1, Minghu2, FChen, JZhao, JQLi}. It is argued that the mesoscopic coexistence with the Fe-vacancy ordered phase K$_2$Fe$_{4}$Se$_5$ could give rise to superconductivity in KFe$_{2}$Se$_2$ \cite{ChenXi}.  However, the phase separation makes it intricate to experimentally determine the intrinsic nature of the superconducting phase . It is crucial to study more HEDIS materials, especially those with decent stability in air and without phase separation,  for understanding the  pairing symmetry and high $T_c$ in HEDIS.

\begin{figure}
\includegraphics[width=8.7cm]{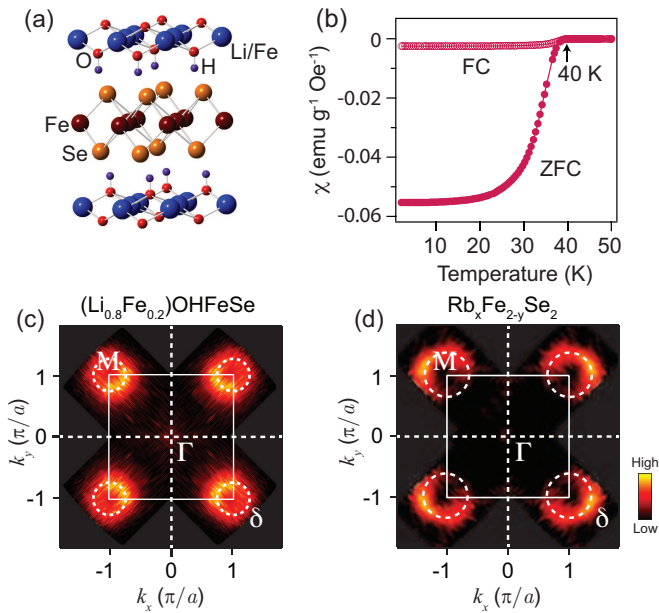}
\caption{(a)The crystal structure and (b) magnetic susceptibility of (Li$_{0.8}$Fe$_{0.2}$)OHFeSe. (c, d) False color plots of the photoemission intensity maps at the Fermi energy ($E_F$) of (Li$_{0.8}$Fe$_{0.2}$)OHFeSe and Rb$_{0.76}$Fe$_{1.87}$Se$_2$ respectively. The intensity was integrated over an energy window of ($E_F$-15 meV, $E_F$+15 meV). The Fermi surface maps are four-fold symmetrized plotted over the projected two-dimensional Brillouin zone for the unit cell of two iron ions. The Fermi surfaces of the $\delta$ band are shown by the dashed circles. In this figure and hereafter, the data of (Li$_{0.8}$Fe$_{0.2}$)OHFeSe and Rb$_{0.76}$Fe$_{1.87}$Se$_2$  were taken with 21.2 eV photons from an in-house helium discharge lamp and 64 eV photons at Beamline I05 of Diamond Light Source, respectively. Taking the inner potential of 11 eV \cite{YZhangMat} to calculate the $k_z$'s of Rb$_{0.76}$Fe$_{1.87}$Se$_2$, we measured at the $\varGamma$-$M$ plane of Rb$_{0.76}$Fe$_{1.87}$Se$_2$. The Fermi surface of (Li$_{0.8}$Fe$_{0.2}$)OHFeSe shows negligible $k_z$ dependence in photon-energy-dependent ARPES study. The data were taken at \textit{T}$<$15 K.}
\label{FS}
\end{figure}

Recently, a new intercalated FeSe-derived superconductor with $T_c$ higher than 40 K, (Li$_{0.8}$Fe$_{0.2}$)OHFeSe, has been synthesized \cite{XFLu}. 
 The crystal structure consists of alternating stacking layers of FeSe and (Li$_{0.8}$Fe$_{0.2}$)OH [Fig.~\ref{FS}(a)]. 
Based on the chemical formula, there are 0.2 electrons per formula unit in the (Li$_{0.8}$Fe$_{0.2}$)OH layer, which could be transferred to the FeSe layer, making this material a new HEDIS candidate, while experiments on its electronic structure are still lacking. 
Moreover, this material shows a good stability in air, and its single crystals can be grown readily with size of millimeters  \cite{ZXZhao}. 
In particular, there is no Fe-vacancy ordered phase or antiferromagnetism in the FeSe layers, which makes this material a promising prototype for studying the superconductivity of HEDIS.

Using angle-resolved photoemission spectroscopy (ARPES), we studied the surface electronic structure of (Li$_{0.8}$Fe$_{0.2}$)OHFeSe. We found that the Fermi surface consists of only electron pockets. The top FeSe layer probed by ARPES shows a carrier concentration of 0.1e$^-$ per Fe, which should be half of the bulk doping level considering the polar discontinuity at the topmost FeSe layer, confirming that (Li$_{0.8}$Fe$_{0.2}$)OHFeSe is an HEDIS material.  Nearly isotropic superconducting gap has been observed around the electron pockets.
 Considering the  chemical potential shift by the additional doping in bulk, our data suggest that the electronic structure and quasiparticle dynamics in (Li$_{0.8}$Fe$_{0.2}$)OHFeSe  are similar to those in $A_x$Fe$_{2-y}$Se$_2$.
  Comparing (Li$_{0.8}$Fe$_{0.2}$)OHFeSe with Rb$_{0.76}$Fe$_{1.87}$Se$_2$, we argue that the crystalline homogeneity of FeSe layers and some unknown effects of (Li$_{0.8}$Fe$_{0.2}$)OH layers in (Li$_{0.8}$Fe$_{0.2}$)OHFeSe could be the key factors that promote its higher $T_c$.

\begin{figure*}
\includegraphics[width=16cm]{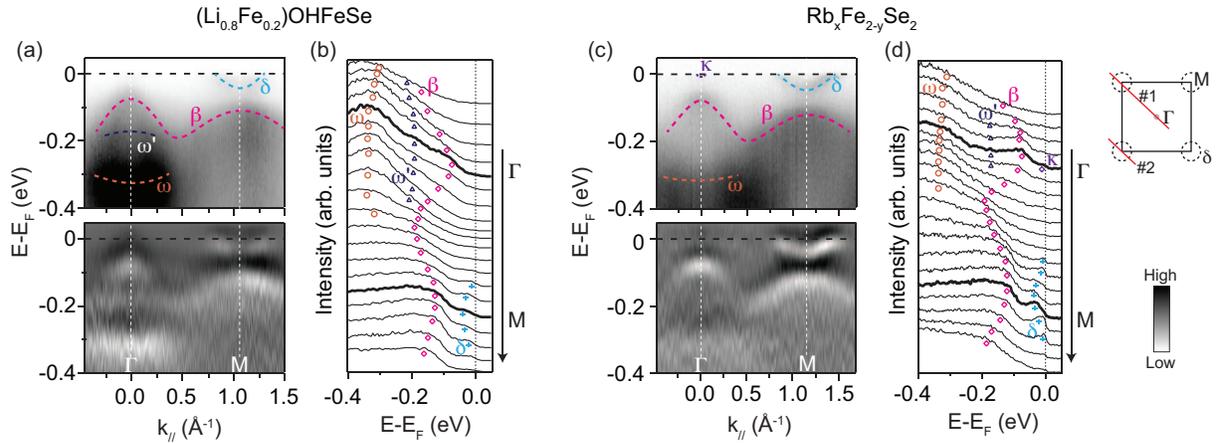}
\caption{(a) Photoemission intensity $I$ ($k$, $\omega$)  (upper panel) of (Li$_{0.8}$Fe$_{0.2}$)OHFeSe and its second derivative  [$\partial$ $_{\omega}^{2}$ $I$ ($k$, $\omega$)] (lower panel) along the $\varGamma$-$M$ direction (cut \#1)  in the right inset. The data were measured at \textit{T}=14 K. (b) The EDCs for the photoemission intensity in panel (a). (c), (d) Same as in panels (a), (b), but the data were taken from Rb$_{0.76}$Fe$_{1.87}$Se$_2$ at \textit{T}=6 K.} \label{band}
\end{figure*}

\begin{figure}[t]
\includegraphics[width=8.7cm]{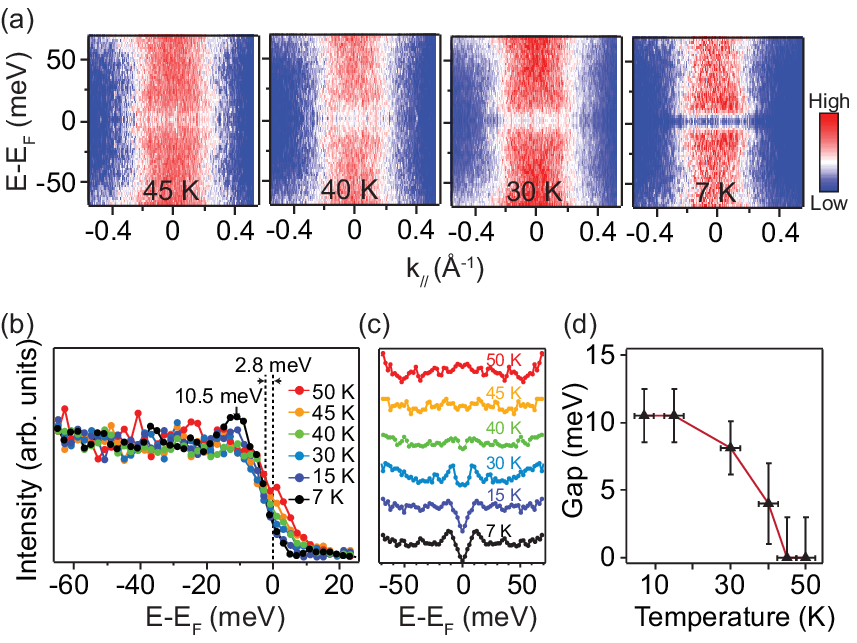}
\caption{(a) The symmetrized (with respect to $E_F$) photoemission intensity of (Li$_{0.8}$Fe$_{0.2}$)OHFeSe at \textit{T}=45,  40, 30, and 7 K along the cut \#2 in the right inset of Fig.~\ref{band}(c). At \textit{T}=40, 30, and 7 K, the spectral weight suppression at $k_F$ near $E_F$ indicates a gap opening. (b) Temperature dependence of the spectral weight integrated over the small momentum region around $k_F$. (c) The spectral weight in  (b) is symmetried with respect to $E_F$ to show the variation of gap size. (d) Temperature dependence of the gap size extracted from (c).}
\label{gapT}  
\end{figure}

Single crystals were grown using the method as described in ref. \cite{XFLu, ZXZhao}. Fig.~\ref{FS}(b) shows  the magnetic susceptibility measurement on a typical (Li$_{0.8}$Fe$_{0.2}$)OHFeSe sample, indicating that the superconductivity occurs at $T_c$ $\sim$ 40 K. 
High-resolution ARPES measurements were performed at the I05 beamline of the Diamond Light Source and our in-house ARPES system with a Helium discharged lamp (21.2 eV photons), using Scienta R4000 electron analyzers. The overall energy resolution was 5$ \sim $12 meV, and the angular resolution was 0.3 degrees. All samples were cleaved \textit{in-situ}  under ultra-high vacuum conditions. During measurements, the spectroscopy qualities were carefully monitored to avoid the sample aging issue.

The single crystals of (Li$_{0.8}$Fe$_{0.2}$)OHFeSe can be easily cleaved along the interface between (Li$_{0.8}$Fe$_{0.2}$)OH layers and FeSe layers due to the weak hydrogen bonding interaction between them  \cite{XFLu}. Patches of both  (Li$_{0.8}$Fe$_{0.2}$)OH and FeSe exist on the exposed surface after cleavage, which has been confirmed by our STM measurements \cite{STM}. Figure \ref{FS}(c) is the photoemission intensity map at Fermi energy ($E_F$) taken using photon energy of 21.2 eV, and negligible $k_z$ dependence of Fermi surface has been confirmed by our photon-energy-dependent ARPES study. The Fermi surface of (Li$_{0.8}$Fe$_{0.2}$)OHFeSe consists of electron pockets around the zone corner \textit{M} and negligible spectral weight at the zone center $\varGamma$.  This Fermi surface topology is identical to that of $A_x$Fe$_{2-y}$Se$_2$ at $\varGamma$-$M$ plane, as exemplified by the photoemission intensity map at $E_F$ in Rb$_{0.76}$Fe$_{1.87}$Se$_2$ shown in Fig.~\ref{FS}(d). Quantitatively, the size of the electron pockets is much smaller in (Li$_{0.8}$Fe$_{0.2}$)OHFeSe than in Rb$_{0.76}$Fe$_{1.87}$Se$_2$. Based on the Luttinger volume, the estimated carrier concentration is 0.1 electrons per Fe in (Li$_{0.8}$Fe$_{0.2}$)OHFeSe, while that of Rb$_{0.76}$Fe$_{1.87}$Se$_2$ is 0.2 electrons per Fe, considering the degenerated $\delta$ and $\delta$' electron pockets at $E_F$. We note that both FeSe and (Li$_{0.8}$Fe$_{0.2}$)OH layers are metallic based on our STM results \cite{STM}, and thus they could both contribute to the spectral weight in ARPES measurements. Nevertheless, our data show the spectral weight only from the FeSe layer, without evident components from (Li$_{0.8}$Fe$_{0.2}$)OH. One possible reason is that the matrix element of the (Li$_{0.8}$Fe$_{0.2}$)OH states is much weaker than that of the FeSe states in our experimental setup.

Due to the surface-sensitive characteristic of ARPES, what we actually measured is the surface electronic structure of (Li$_{0.8}$Fe$_{0.2}$)OHFeSe, which should be different from the bulk case because of the polar discontinuity at the surface. The top FeSe layer has only one adjacent charge reservoir layer of (Li$_{0.8}$Fe$_{0.2}$)OH, and compared with the FeSe layers in the bulk, it retains only half of the doped charges based on simple polar analysis \cite{LSMO}. Nevertheless, with the preserved in-plane structure parameter/bond length, the variation of electron carriers on the top FeSe layer would mostly affect the chemical potential other than the band structure or the electron correlation, as demonstrated in Ba$_{1-x}$K$_x$Fe$_2$As$_2$ with different K doping \cite{ZRYe}. Based on this, we argue that the bulk Fermi surface of (Li$_{0.8}$Fe$_{0.2}$)OHFeSe could be similar to that in Rb$_{0.76}$Fe$_{1.87}$Se$_2$, in both topology and pocket size, and the surface electronic structure measured by ARPES could to some extent reveal the quasiparticle dynamics and electron correlation of bulk (Li$_{0.8}$Fe$_{0.2}$)OHFeSe.

The surface band dispersions of (Li$_{0.8}$Fe$_{0.2}$)OHFeSe are shown in Fig.~\ref{band}. Figures~\ref{band} (a),(c) show the photoemission intensity  along the cut \#1 together with the second-derivative images for (Li$_{0.8}$Fe$_{0.2}$)OHFeSe and Rb$_{0.76}$Fe$_{1.87}$Se$_2$, respectively. The  corresponding energy distribution curves (EDCs) are plotted in Figs.~\ref{band} (b),(d). The band structures of (Li$_{0.8}$Fe$_{0.2}$)OHFeSe and Rb$_{0.76}$Fe$_{1.87}$Se$_2$ are rather similar. We are able to resolve a dispersive $\beta$ band under the $E_F$ and an electron-like band $\delta$ around $M$ point. Around 200~meV  and 330~meV below $E_F$, we observe two rather flat $d_{z^2}$ bands assigned as $\omega$' and $\omega$ respectively in (Li$_{0.8}$Fe$_{0.2}$)OHFeSe, which are weak but still present in  Rb$_{0.76}$Fe$_{1.87}$Se$_2$, and have also been resolved in the previous ARPES results on superconducting K$_{x}$Fe$_{2-y}$Se$_2$ \cite{FChen_orb}.  In particular, the bandwidths of the $\beta$ are nearly the same for these two compounds, suggesting that the band dispersions and correlation strength of (Li$_{0.8}$Fe$_{0.2}$)OHFeSe is very similar to those of $A_x$Fe$_{2-y}$Se$_2$.  We also note that there are some spectra weight at $E_F$ around $\varGamma$ point in Rb$_{0.76}$Fe$_{1.87}$Se$_2$ due to the strong $k_z$ dispersion of $\kappa$  around $Z$ point of the Brillouin zone \cite{YZhangMat, FChen_orb}, which is absent in the spectra of (Li$_{0.8}$Fe$_{0.2}$)OHFeSe. This discrepancy could be accounted by the lighter electron doping in the surface layer of (Li$_{0.8}$Fe$_{0.2}$)OHFeSe.

\begin{figure}[t]
\includegraphics[width=8.7cm]{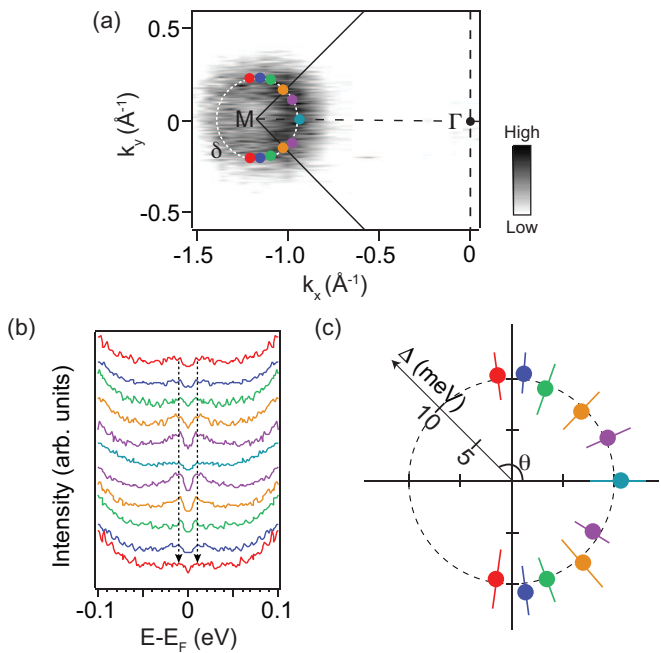}
\caption{(a) Photoemission intensity map of the $\delta$ band around the zone corner \textit{M}. The data were integrated over an energy window of ($E_F$-15 meV, $E_F$+15 meV). (b) The symmetrized (with respect to E$_F$) EDCs taken at various Fermi crossings at the $\delta$ band, as indicated by the dots in (a). (c) Gap distribution of the $\delta$ band around \textit{M} in polar coordinates, where the radius indicates the gap size, and the polar angle $\theta$ represents the angle on the $\delta$ band with respect to \textit{M}, with $\theta$=0 being the $\varGamma$-\textit{M} direction.}
\label{gapM}
\end{figure}

The evolution of the superconducting gap with temperature is plotted in Fig.~\ref{gapT}. As shown in Fig.~\ref{gapT}(a), there is no gap at $E_F$ near the Fermi momentum in the normal state, whereas a clear superconducting gap shows up at low temperatures in the superconducting state. With decreasing temperature, the spectra weight near the $E_F$ is depleted and a coherent peak grows in Fig.~\ref{gapT}(b). We identify that the value of leading edge gap is  $ \sim $ 2.8 meV. Using the coherent peak position as a measurement of superconducting gap, we estimate that it is $ \sim $10.5 meV, very similar to the value in $A_x$Fe$_{2-y}$Se$_2$ materials \cite{YZhangMat}. In Fig.~\ref{gapT}(c), the symmetrized EDCs with respect to $E_F$ demonstrate the gap opening with decreasing temperature. Fig.~\ref{gapT}(d) shows the superconducting gap size as a function of temperature determined from Fig.~\ref{gapT}(c), clearly indicating a $T_c$ around 40 K, which is consistent with the results of the magnetic susceptibility measurements (Fig.~\ref{FS}(b)). By examining the superconducting gap size at various Fermi crossings of the $\delta$ band around $M$, we plot its momentum distribution in Fig.~\ref{gapM}. Within the experimental uncertainties, the superconducting gap is isotropic with a size of $\sim$ 10 meV. The gap amplitude is similar to that of $A_x$Fe$_{2-y}$Se$_2$ \cite{YZhangMat}. The $T_c$ of (Li$_{0.8}$Fe$_{0.2}$)OHFeSe  is 30\% higher than that of Rb$_{0.76}$Fe$_{1.87}$Se$_2$, so we naturally expect that the former has a superconducting gap greater than 10 meV. The observation of 10 meV gap size here could be due to the fact that the top FeSe surface layer is underdoped compared with the FeSe layers in the bulk. Due to the proximity effect, this underdoped surface FeSe layer shows a smaller gap value but a gap-closing temperature close to the bulk  $T_c$. Other bulk sensitive experimental probes are required to reveal the bulk superconducting gap  in  (Li$_{0.8}$Fe$_{0.2}$)OHFeSe.

The FeSe layers are the key structural elements in intercalated FeSe-derived superconductors. We have shown that the electronic structure of FeSe layers in (Li$_{0.8}$Fe$_{0.2}$)OHFeSe highly resembles what have been observed in $A_x$Fe$_{2-y}$Se$_2$. One may wonder why the former has a higher $T_c$ than the latter. One possibility is that an intact FeSe layer, without an intertwined Fe-vacancy order or antiferromagnetic phase, may naturally bear a $T_c$ higher than 40 K. When there is inhomogeneity and intergrowth of multiple phases, the $T_c$ will be suppressed. Such a scenario may find evidence in some $A_x$Fe$_{2-y}$Se$_2$ compounds, though the bulk superconducting transition occurs well below 40 K, there is evidence showing some tiny superconducting phase with  a $T_c$ slightly above 40 K \cite{KFeSe40K1, KFeSe40K2}. This particular sign of higher $T_c$ can be found in samples with large domains that seem to retain intact FeSe layers without Fe vacancies \cite{KFeSe40K2}. Another possible reason is that (Li$_{0.8}$Fe$_{0.2}$)OH layer may help to enhance the $T_c$ in the FeSe layers. In analogy with the interface superconductivity in the electronic system of monolayer FeSe film on an SrTiO$_3$ substrate, where the substrate dopes electrons into the FeSe layer and some phonon mode likely boosts the $T_c$ \cite{ZXShen}, it is very intriguing to investigate whether similar physics may occur in (Li$_{0.8}$Fe$_{0.2}$)OHFeSe with relatively high $T_c$. However, phonon-induced replica bands observed in single layer FeSe/STO is not observed here, indicating the electron-phonon interaction strength is much weaker in (Li$_{0.8}$Fe$_{0.2}$)OHFeSe. Moreover, as a spacing layer, (Li$_{0.8}$Fe$_{0.2}$)OH layer may weaken the interlayer coupling between neighboring FeSe layers, which can make the electronic system in FeSe layers more two-dimensional with amplified electronic, magnetic or orbital fluctuation and lead to higher $T_c$. More studies are demanded for uncovering the physics of superconductivity in (Li$_{0.8}$Fe$_{0.2}$)OHFeSe as well as other intercalated FeSe-derived superconductors.

In summary, we have studied the surface electronic structure of  intercalated FeSe-derived superconductor (Li$_{0.8}$Fe$_{0.2}$)OHFeSe, where there is no Fe-vacancy order or antiferromagnetism in the FeSe layer. Without evidence of other intertwined phase,  (Li$_{0.8}$Fe$_{0.2}$)OHFeSe offers an excellent platform to study intercalated FeSe-derived superconductors. We found that (Li$_{0.8}$Fe$_{0.2}$)OHFeSe has a electronic structure akin to that of Rb$_{0.76}$Fe$_{1.87}$Se$_2$, with similar band dispersions and gap symmetry. The higher $T_c$ in (Li$_{0.8}$Fe$_{0.2}$)OHFeSe could be due to the superior quality of FeSe layer in this material. The (Li$_{0.8}$Fe$_{0.2}$)OH layer may also play an important role for elevating $T_c$. Further investigations are required to reveal what is the critical factor that enhances the $T_c$.

Some of the preliminary data (not shown here) were taken at the beamline 21B1 of National Synchrotron Radiation Research Center(NSRRC), beamline 13U of National Synchrotron Radiation Laboratory (NSRL), and the beamline 5-4 of the Stanford Synchrotron Radiation Light source (SSRL). We gratefully acknowledge the experimental support by Dr. M. Hoesch, Dr. T. Kim, and Dr. P. Dudin at Diamond Light Source, Dr. C. M. Cheng and Prof. K. D. Tsuei at NSRRC, Dr. D. H. Lu and Dr. H. Makoto at SSRL, as well as helpful discussions with J. P. Hu. This work is supported by the National Science Foundation of China and National Basic Research Program of China (973 Program) under Grants No. 2012CB921402, No. 2012CB927401, No. 2011CB921802, No. 2011CBA00112, and No. 2011CBA00106. SSRL is operated by the US DOE Office of Basic Energy Science.

Note: upon finishing this work, we noticed another independent ARPES study on (Li$_{0.8}$Fe$_{0.2}$)OHFeSe, which revealed similar band structure and superconducting gap distribution, by Lin Zhao $et$ $al$ posted in arXiv: 1505.06361.


\begin{thebibliography}{10}



\bibitem{XLChen} J. G Guo, S. F Jin, G. Wang, S. C. Wang, K. X. Zhu, T. T. Zhou, M. He, and X. L. Chen, Phys. Rev. B \textbf{82}, 180520(R) (2010).

\bibitem{KFeSe40K1} T. P. Ying, X. L. Chen, G. Wang, S. F. Jin, T. T. Zhou, X. F. Lai, H. Zhang, W. Y. Wang, Sci. Rep., \textbf{2}, 426 (2012).



\bibitem{QKXue} Q. Y. Wang $et$ $al$., Chin. Phys. Lett. \textbf{29}, 037402 (2012).

\bibitem{XJZhou1} Defa Liu $et$ $al$., Nat. Comm. \textbf{3}, 931 (2012).

\bibitem{XJZhou2}Shaolong He $et$ $al$., Nat. Mater.  \textbf{12}, 605 (2013).


\bibitem{SYTan} S. Y. Tan $et$ $al$., Nat. Mater. \textbf{12}, 634 (2013).

\bibitem{RPeng}R. Peng $et$ $al$., Phys. Rev. Lett. \textbf{112}, 107001 (2014).

\bibitem{RPengN} R. Peng $et$ $al$., Nat. Comm. \textbf{5}, 5044 (2014).

\bibitem{YZhangMat} Y. Zhang $et$ $al$., Nat. Mater. \textbf{10}, 273 (2011).

\bibitem{Hirschfeld} P. J. Hirschfeld,  $et$ $al$., Reports On Progress In Physics, \textbf{74}, 124508 (2011).


\bibitem{JFJia} J. F. Ge,	Z. L. Liu,	C. H. Liu, C. L. Gao,	D. Qian,	Q. K. Xue, Y. Liu, and J. F. Jia, Nat. Mater.  \textbf{14}, 285 (2015).
\bibitem{ZXShen} J. J. Lee $et$ $al$., Nature, \textbf{515}, 245 (2014).

\bibitem{Minghu1} M. H. Fang $et$ $al$., Europhys. Lett. \textbf{94}, 27009  (2011).
\bibitem{Minghu2} H. D. Wang $et$ $al$., Europhys. Lett. \textbf{93}, 47004 (2011). 
\bibitem{FChen}  F. Chen $et$ $al$., Phys. Rev. X \textbf{1}, 021020 (2011).
\bibitem{JZhao} J. Zhao, H. Cao, E. B-Courchesne, D. -H. Lee, R. J. Birgeneau, Phys. Rev. Lett. \textbf{109}, 267003 (2012).
\bibitem{JQLi} Z. Wang $et$ $al$., Phys. Rev. B \textbf{83}, 140505 (R) (2011). 
\bibitem{ChenXi} Wei Li $et$ $al$., Phys. Rev. Lett. \textbf{109}, 057003 (2012).
\bibitem{XFLu} X. F. Lu $et$ $al$., Nat. Mater. \textbf{14}, 325 (2015).
\bibitem{ZXZhao} X. L. Dong $et$ $al$., arXiv: 1502.04688.

\bibitem{STM} Y. J. Yan $et$ $al$. (unpublished)

\bibitem{LSMO} R. Peng,  $et$ $al$., Applied Physics Letters, \textbf{104}, 081606 (2014).

\bibitem{ZRYe} Z. R. Ye $et$ $al$., Phys. Rev. X \textbf{4}, 031041 (2014).

\bibitem{FChen_orb} F. Chen, Q. Q. Ge, M. Xu, Y. Zhang, X. P. Shen, W. Li, M. Matsunami, S.-i. Kimua, J. P. Hu, and D. L. Feng, Chin. Sci. Bull. \textbf{57}, 3829 (2012).



\bibitem{KFeSe40K2}Masashi Tanaka $et$ $al$., arXiv: 1504.04197.






\end{thebibliography}
\end{document}